\documentclass[12pt]{iopart}
\usepackage{epsfig}

\begin{document}

\title[Neutrino Lensing]{Neutrino Lensing}

\author[]{
LUO Xin-Lian ({}\hspace{20 mm})}

\address{
{$^1$ Department of Astronomy, Nanjing University, Nanjing, 210093,
China}}

\begin{abstract}
Due to the intrinsic properties of neutrinos, the gravitational lens
effect for neutrino should be more colorful and meaningful than the
normal lens effect of photon. Other than the oscillation experiments
operated at terrestrial laboratory, in principle, we can propose a
completely new astrophysical method to determine not only the nature
of gravity and spacetime of lens objects but also the mixing
parameters of neutrinos by analyzing neutrino trajectories near the
central objects. However, compared with the contemporaneous
telescopes through the observation of the electromagnetic radiation,
the angular, energy and time resolution of the neutrino telescopes
are still comparatively poor, we just concentrate on the two
classical tests of general relativity, i.e. the angular deflection
and time delay of neutrino by a lens object as a preparative work in
this paper. In addition, some simple properties of neutrino lensing
are investigated.

\end{abstract}

\ead{\mailto{xlluo@nju.edu.cn}}

\vspace{.3in}


\noindent Neutrinos are one kind of the fundamental particles that
make up the universe. Up to date, many interesting neutrino
astrophysical sources have been known, such as the ultra-low energy
Big Bang relic background neutrinos, the $\sim 10$ keV  stellar
neutrinos, the $\sim 10$ MeV supernova neutrinos, the $\sim 100$ TeV
neutrinos from young supernova remnants in our galaxy, the
ultra-high energy neutrinos from Active Galactic Nucles (AGNs),
Gamma Ray Bursts (GRBs), Microquasars or some other celestial
sources. Since neutrinos are electrically neutral leptons, they only
participate in the weak interaction and are almost unaffected
through transmission, and they open a new window into space to the
high-energy processes in the universe to astronomy. As we known,
several under-ice and under-water neutrino telescopes are being or
scheduled to be constructed now, such as Icecube located at the
south Pole worked in energy range from GeV to PeV, expected to
achieve an angular resolution below $1^{\circ}$, a real-time
resolution at $2$\,ns and an energy resolution below $0.3$ in $\log
(E)$.

In addition, neutrinos are also hot topics and research areas for
particle physicists. One of the most remarkable discoveries in
particle physics connect with neutrinos in the last few decades is
the finding that neutrinos are massive and mixed particles. Although
the absolute scale of neutrino masses is still unknown, the evidence
is strong. The precision measurement of neutrino masses and
oscillation parameters now is one of the main goals of the neutrino
experiments. It is possible for researchers to measure the absolute
mass of neutrino with an accuracy of $0.05$\,eV soon by the double
beta decay experiment in many laboratories or by the lensing of the
Cosmic Microwave Background radiation (CMB) from cosmological
probes.

As we known, a light ray can be deflected by gravity, it is same for
neutrino. Since Soldner firstly published the derivation of the
deflection of light by a massive object based on the framework of
Newtonian gravity almost two centuries ago,$^{\cite{Soldner}}$ the
gravitational lensing has been full investigated numerically or
analytically. Now, lensing of light ray has been used widely in the
research of astrophysics and already became an excellently
educational stuff for students of astronomy.$^{\cite{Schneider}}$
Another kind of gravitational lens researchers less mentioned is
neutrino lensing, which has also been put forward and widely studied
for a quite long time. Unlike photon, neutrino can penetrate into
most of celestial objects easily except black holes or some special
compact objects since it only participates in the gravitational and
weak interaction. Therefore, one obvious advantage of neutrino
lensing in principle is that it may provide a distinct method to
probe the mass distribution of lens objects. Based on the standard
solar model, Gerver investigated the possibility of neutrino
focusing by the Sun's core.$^{\cite{Gerver}}$ Escribano {\it et
al.}$^{\cite{Escribano}}$ discussed the gravitational lensing of
neutrinos by some extended sources as stars, galaxies and galactic
halos. Moreover, Mena {\it et al.}$^{\cite{Mena}}$ studied the
lensing effect of supernova neutrinos by the center black hole of
our Galaxy. Eiroa and Romero$^{\cite{Eiroa}}$ analyzed the lensing
effect of the cosmological sources of neutrinos by some supermassive
black holes. However, all of works cited here used the lightray
approximation, treated neutrino as a massless particle. Strictly
speaking, neutrino does not move along null geodesics since it has a
non-zero rest mass. Based on this idea, Barrow and
Subramanian$^{\cite{Barrow}}$ offered a possible explanation for the
time delay of neutrinos from supernova SN1987a.

In this Letter, we will show that this small but non-zero rest
masses of neutrino will bring us some tinily small but non-trivial
difference of the angular deflection and time delay of neutrino in
gravitational field. In principle, neutrino lensing itself may
provide a unique ground for probing the nature of neutrinos. Thus,
we investigate gravitational lensing of neutrinos by the point-mass
lens and re-investigate two famous effects of General Relativity,
i.e. the deflection and the Shapiro time delay, we extend them to
more general case.

We start from the Kerr metric in Boyer-Lindquist (BL) coordinates
\begin{equation}\label{1}
ds^{2}=\rho ^{2} (\frac{dr^{2}}{\Delta}+d \theta^{2})+(r^{2}+a^{2})
\sin^{2}\theta d \phi^{2} -c^{2} dt^{2}+\frac{r_{s} r}{\rho^{2}} (a
\sin^{2}\theta d \phi - c dt)^{2}
 \; ,
\end{equation}
where $r_{s} = 2 G M / c^{2}$ is the Schwarzschild radius, $a=J / M
c$ is the specific angular momentum with the dimension of length,
$\rho^{2}=r^{2}+a^{2} \cos^{2} \theta$ and $\Delta = r^{2}-r_s
r+a^{2} $. Since this metric describe a stationary and axisymmetric
spacetime,  it is easy for us to obtain those conserved quantities
for test particles in Kerr geometry,
\begin{equation}\label{2}
E/mc =   (1-\frac{r_{s} r}{\rho^{2}})c  \dot{t}+\frac{r_{s}
r}{\rho^{2}} a \sin^{2}\theta \dot{\phi}  \;,
\end{equation}
\begin{equation}\label{3}
L_{z} /m =  -\frac{r_{s} r}{\rho^{2}} a \sin^{2}\theta c \dot{t} +
\frac{(r^{2}+a^{2})^{2}-\Delta a^{2} \sin^{2}\theta}{\rho^{2}}
\sin^{2}\theta \dot{\phi} \;.
\end{equation}
The dots denote differentiation with respect to an affine parameter.
The constants $E/m$ and $L_{z}/m$ can be viewed as the energy and
the axial component of angular momentum per unit mass at infinity.
In addition, other two constants of motion are the particle's rest
mass $m$,
\begin{equation}\label{4}
mc= |p| =\left( - g^{\mu \nu} p_{\mu} p_{\nu}\right)^{1/2}\;,
\end{equation}
and Carter's constant $Q$,
\begin{equation}\label{5}
Q =p_{\theta}^{2}+\cos^{2}\theta \left[  a^{2}
(m^{2}c^{2}-E^{2}/c^{2})+\left( L/\sin \theta \right)^{2}\right] \;.
\end{equation}
For simplicity, we only investigate the motion for the open (or
infinite) orbits in the equatorial plane, i.e. $E \geq mc^{2}$,
$\theta = \pi /2$ and $p_{\theta} = 0$. Considering an incoming
particles with Lorentz factor $\gamma=E/mc^{2}$and impact parameter
$b\approx  L_{z} / (\sqrt{\gamma^2-1} mc)$ at distance, we can solve
for the velocity components in terms of $\gamma$ and $b$
\begin{equation}\label{6}
\dot t = \frac{1}{{\Delta }}\left[ {\left( {r^2  + a^2  + \frac{{a^2
r_s }}{r}} \right) \gamma - \frac{{ar_s }}{r} b{\sqrt{\gamma^2-1}}}
\right] \;,
\end{equation}
\begin{equation}\label{7}
\dot \phi  = \frac{c}{{\Delta }}\left[ { \frac{{ar_s }}{r}\gamma}
+\left( {1 - \frac{{r_s }}{r}} \right) b{\sqrt{\gamma^2-1}}\right]
\;,
\end{equation}
\begin{equation}\label{8}
\dot r^2  = \frac{c^{2}}{{r^3 }}\left[ {\gamma^2 r^3  + r_s \left(
{a\gamma - b{\sqrt{\gamma^2-1}} } \right)^2  + r\left( {a^2
\gamma^{2} - b^2 \gamma^{2}+b^{2}} \right)} \right] - \frac{{c^2
\Delta }}{{r^2 }}  \;.
\end{equation}
It is obvious that the above equations are not only suitable for
null geodesics for massless particles like the photon ($\gamma \to
\infty$) but also suitable for time-like geodesics for massive
particles with finite $\gamma$. The orbit of a photon is more simple
since it is not decided by its energy but completely decided by its
impact parameter. We will focus on the general behavior of test
particles (including photons) hereafter.

Setting $dr/d\phi=0$, we can derive a relationship between the
turning point (the point of closest approach to central object)
$r_{0}$ and the impact parameter $b$,
\begin{equation}\label{9}
b_{\pm} = \left[\sqrt{\gamma^2-1}(1-\frac{r_{s}}{r_{0}})
\right]^{-1} \left[ -a \gamma \frac{r_{s}}{r_{0}} \pm  r_{0}
\sqrt{(1-\frac{r_{s}}{r_{0}}+\frac{a^{2}}{r_{0}^{2}})
(\gamma^{2}-1+\frac{r_{s}}{r_{0}})} \; \right] \;,
\end{equation}
where $+$ and $-$ denote prograde and retrograde particles
respectively. We also can express $r_{0}$ as a power series in $u =
r_{s}/|b|$,
\begin{eqnarray}\label{10}
r_{0}/|b| & = &  1 -\frac{1+s^2}{2 s^2} u -
\left(\frac{3}{8}+\frac{1}{4 s^2}-\frac{1}{8
   s^4} -\frac{\tilde{a} \sqrt{1+s^2}}{s}+\frac{\tilde{a}^2}{2}\right) u^2 \nonumber \\
& & -\left[\frac{1}{2}+\frac{1}{2 s^2}-\frac{\tilde{a} \sqrt{1+s^2}
   \left(1+3 s^2\right)}{2 s^3}+\frac{1}{2} \tilde{a}^2
\left(2+\frac{1}{s^2}\right)\right] u^3 \nonumber \\
& & -\left[\frac{105 s^8+140 s^6+30 s^4-4 s^2+1}{128
s^8}-\frac{\tilde{a} \sqrt{1+s^2} \left(2+3
   s^2\right)}{s^3} \right. \nonumber \\
& & \left. +\frac{\tilde{a}^2 \left(51 s^4+46
   s^2+3\right)}{16 s^4}-\frac{\tilde{a}^3 \sqrt{1+s^2}}{s}+ \frac{\tilde{a}^4}{8} \right] u^{4}
+{\cal{O}}(u^5)\;,
\end{eqnarray}
where $s = \sqrt{\gamma^2-1}$, $\tilde{a} = a / r_{s}$ is the
specific angular momentum in units of the Schwarzschild radius, with
$\tilde{a} > 0$ for prograde particles and $ \tilde{a} < 0$ for
retrograde particles in a unified expression.

Simply and repeatedly used the mathematical relation $\partial_{u}
\left[ \int_0^{f(u)} g(x,u) \, dx \right] = \int_0^{f(u)}
\partial_{u}[g(x,u)] \, dx + f'(u)\, g(f(u),u)$, the deflection
angle (in radians) can be expressed as a Taylor series about $u$ as
\begin{eqnarray}\label{11}
\delta \phi & = &2\int\limits_{r_0 }^\infty  {\left( {{\dot \phi
}}/{{\dot r}} \right)dr}  - \pi \nonumber \\
& = & \left(2 + \frac{1}{s^2} \right) u +\left[\frac{3}{16} \, \pi (
5 + \frac{4}{s^2} )- \frac{2 \tilde{a} \sqrt{1+s^2}}{s} \right]
u^{2} + \frac{1}{6} \,
\left[\left(32+\frac{36}{s^2}+\frac{6}{s^4}-\frac{1}
   {2s^6}\right) \right. \nonumber \\
&  & \left. - \frac{\pi \tilde{a}
   \sqrt{1+s^2} \,(2+5 s^2)}{2 s^3}+\left(2+\frac{1}{s^2}\right)\tilde{a}^2 \right]
u^3+\left[ \frac{105 \, \pi}{1024} \,
   \left(33+\frac{48}{s^2}+\frac{16}{s^4}\right) \right. \nonumber \\
&  & \left. -\frac{3 \tilde{a} \sqrt{1+s^2} \,(1+12
   s^2+16 s^4)}{2 s^5}+\frac{3 \pi \tilde{a}^2 (8+88
   s^2+95 s^4)}{64 s^4}-\frac{2 \,\tilde{a}^3 \sqrt{1+s^2}}{s}
\right] u^{4}\nonumber \\
&  & +{\cal{O}}(u^5)\;.
\end{eqnarray}
\begin{figure}[t]
\begin{picture}(190,220)
\put(0,0){\includegraphics{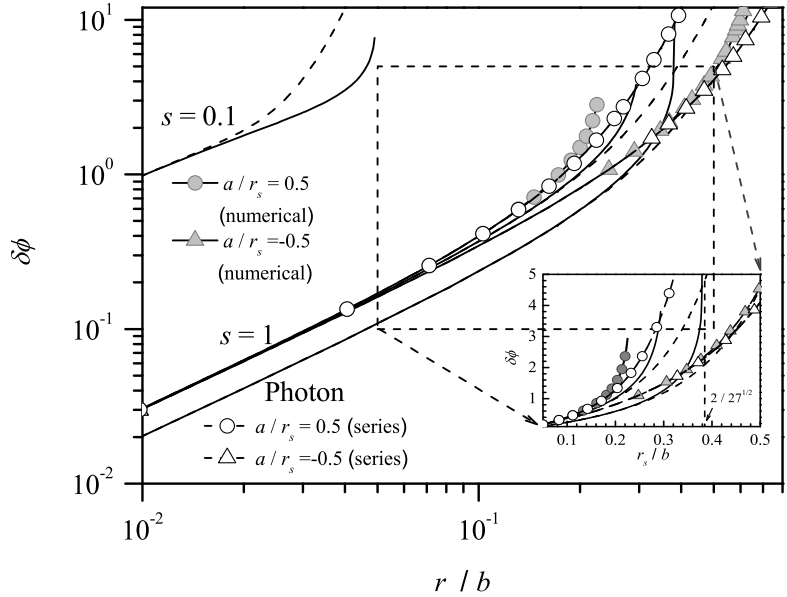}}
\end{picture}
\caption {Deflection angle for photon or some other relativistic
particles with different impact parameters. The solid and the dashed
curves denote the numerical solutions and the series solutions,
respectively. The circle and the triangle curves represent the
deflection of the retrograde and the prograde particles (with $s =
1$ as a demonstration) in the equatorial plane of the extreme Kerr
Black Hole.}
\end{figure}\label{fig1}
If you do not care about the interminable derivation and
integration, definitely you can expand the above integration to any
order you want. As a demonstration, geodesics are expanded as a
Taylor series up to and including fourth-order terms in $r_{s}/|b|$
here. When $a=0$ and $s\to\infty$, we recover the famous first order
light deflection in Schwarzschild space-time, $2 r_{s}/b$, and add
several high order modification. Fig.1 shows the logarithm plot of
the deflection angle for photon and some other relativistic
particles with different impact parameters. For the convenience of
comparison, we have carried out the deflection integral numerically,
i.e. the solid curves in Fig.1. Clearly, the Taylor series expansion
offered here are quite well formula to specify the bending angle
especially for small angle, weak field (with $|b|$ much larger than
$r_{s}$) and ultra-relativistic particles (with $s \gg 1$). The
intermediate relativistic particles with $s = 1$ are taken as an
example to demonstrate the rotational contribution. It is natural
and obvious that the spin is a high-order modification, which only
appears in second-order or even higher order terms of the expansion,
which means that the spin parameter of black hole only can be
determined by the careful observation of the large angle scattering
of the intermediate or ultra relativistic particles. Due to the
frame-dragging effect of the rotating black holes, the particle
traveling in the direction of rotation of the object will move
around the central object faster than particles moving against the
rotation, the position of the turning point for prograde particle
will be even more far away from the center than that of retrograde
particles. Thereupon the deflection angle is enhanced for retrograde
particles (the circle curves) and reduced for prograde particles
(the triangle curves). The vertical dashed line in the infixed plot,
$r_{s}/b = 2 / \sqrt{27}$, corresponding to the critical impact
parameters $b_{c}$ for Schwarzschild black hole. For $b \le b_{c}$,
the photon will spiral in and be captured by the central black hole.
For more general case in Schwarzschild spacetime, the critical value
is
\begin{equation}\label{12}
b_{c} = r_{s} \, \frac{\sqrt{\left(3 \gamma ^2-2\right)^2 \left(3
\gamma ^2-4\right)+\gamma  \sqrt{9 \gamma ^2-8} \left(9 \gamma ^4-20
\gamma ^2+12\right)}}{\sqrt{2} \left(\gamma ^2-1\right)
\left(\sqrt{9 \gamma ^2-8}-\gamma \right)}
 \;.
\end{equation}
However there is no analytic expression for the critical impact
parameter in Kerr black hole. By considering the capture of the
background photons, hot, warm and cold dark matter in the early
Universe, we plan to investigate the growth of primordial black
holes in our future work, which are possible seeds for supermassive
black holes at the center of the galaxies.

We have discussed the angular deflection of photon or relativistic
particle by a rotating spherical body. From the first order term of
Eq.\,(11), we can see that the difference of the deflection between
photon and ultra-relativistic particle, $2 r_{s}/b (\gamma^{2}-1)$,
is tinily small, which means that the neutrino lens and the photon
lens are nearly the same in weak deflection limit, that is why
people can safely and simply use null geodesics as photons do to
study neutrino lensing.$^{\cite{Mena, Eiroa}}$ If we want to
directly determine the mass of neutrinos by bending, we must seek
for the observation of the large angle scattering. Unlike photons,
neutrinos are electrical neutrality, only participate in the weak
interaction, can penetrate the lens objects easily, hence it should
be possible for researchers to detect those large angle scattering
neutrinos in principle. However, it is difficult for us to find the
optical counterpart of the same source to compare with due to the
strong dust extinction in the neighborhood of the lens objects.
Furthermore, the requirement for angular resolution may go far
beyond the capability of neutrino telescopes at present even in the
near future, the operation process to detect the bending angle
precisely will be very difficult and nearly hopeless for us. Thus,
it seems that our work here is more meaningful from the theoretical
than from the observational point of view. In fact, we really plan
to study the neutrino lensing by a Kerr black hole in weak
deflection limit with higher order modification by following
Sereno's work$^{\cite{Sereno}}$ in a forthcoming paper. In the
following paragraphs, we only devote to the study of the properties
of neutrino lensing as the simplest, i.e. the point-mass lens model
is employed. In addition, we only consider the contribution of the
first order term for simplicity.

\begin{figure}
\begin{picture}(190,220)
\put(0,0){\includegraphics{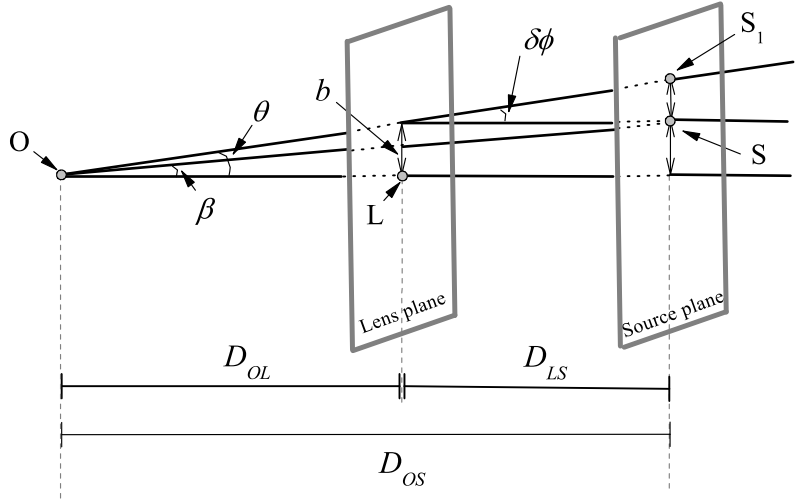}}
\end{picture}
\caption {Sketch of a typical lensing system.}
\end{figure}\label{fig2}
If there are no other deflectors except a lensing object of mass $M$
close to the line-of-sight to a source, the sketch map of a typical
lensing system can be drawn easily (See Fig.\,2). The distances
between source and observer, source and lens, and observer and lens
are given by $D_{OS}$, $D_{LS}$, $D_{OL}$, respectively. The angle
$\beta$ and $\theta$ denote the positions of source and images with
respect to the lens direction. If the physical size of the lensing
object is much smaller than both $D_{LS}$ and $D_{OL}$, and the all
angles ($\beta$, $\theta$, $\delta\phi$) are very small, from
Fig.\,2 we can reach the geometric relation $\beta=\theta-\delta
\phi \, D_{LS}/D_{OS}$. Notice $b = \theta \, D_{OL}$ and the
deflection angle expression to first order, we obtain the famous
lens equation $\beta = \theta - \theta_{E}^{2}/\theta$, where the
Einstein angle $\theta_{E}$ now satisfies
$\theta_{E}^{2}=(2+1/s^{2}) \, r_{s} {D_{LS}}/({D_{OS}D_{OL}})$,
which defines the angular scale for a lens situation. By solving the
lens equation, one can find the positions of the primary (upper
sign) and the secondary (lower sign) images for an isolated point
source, $\theta_{1,2}=\frac{1}{2}\left( \beta \pm \sqrt{\beta^{2}+4
\theta_{E}^{2}} \right)$, which correspond to the shortest geodesic
and the longest geodesic with $b_{1} = \theta_{1} D_{OL} $ and
$b_{2} = |\theta_{2}| D_{OL} $, respectively.

As we known, times can be measured to much greater accuracy than
angles, we will further investigate another prominent and equivalent
effect of General Relativity, i.e. the Shapiro time delay or
gravitational time delay here. This effect was first pointed by
Shapiro in 1964 about the time delay of signals passing near a
massive object.$^{\cite{Shapiro}}$ Let us calculate the time
required for a particle to go form a source point with $r=r_{S}$,
$\phi=\phi_{S}$, to an observer at $r=r_{O}$, $\phi=\phi_{O}$ in the
equatorial plane. The equation governing the time history of orbits
is easy to be given by Eq.\,(6) and (8). The time required for
particle to go from the turning point $r_{0}$ to $r$ is
\begin{eqnarray}\label{13}
t(r,r_{0})& = & \int\limits_{r_0 }^{r}  {\left( {{\dot t }}/{{\dot
r}} \right)dr} \nonumber \\
 & \simeq & \frac{\sqrt{r^2-r_0^2}}{c \left.\sqrt{\gamma ^2-1}\right/\gamma} + \frac{r_{s}}{2
c} \frac{\gamma ^3 }{\left(\gamma ^2-1\right)^{3/2}} \left\{
\sqrt{\frac{r-r_0}{r+r_0}} +  \frac{2 \gamma ^2-3 }{\gamma ^2} \cdot
\ln\left(\frac{r+\sqrt{r^2-r_0^2}}{r_0}\right) \right.
\nonumber \\
 &   &
- \frac{a \left(a-2 a \gamma ^2+2 \gamma  \sqrt{\gamma ^2-1}
\sqrt{a^2+r_0^2}\right)}{\gamma ^2 r_0^2}\sqrt{\frac{r-r_0}{r+r_0}}
- \frac{\left(\gamma ^2-1\right)^{3/2}}{\gamma ^3 } \cdot
\nonumber \\
 &  & \left.
\ln\left[\frac{\left(a^2+r^2\right) r_0^2}{r^2 r_0^2+a^2 \left(2
r^2-r_0^2\right)-2 a r \sqrt{\left(r^2-r_0^2\right)
\left(a^2+r_0^2\right)}}\right]   \right\} \;.
\end{eqnarray}
We only keep the first order in $r_{s}/r$ in the integrand function.
The result show that the traveled the  time has two different
components. The leading term is related to the geometrical of the
spacetime in the absence of the lens, now is the time for the
relativistic particle traveled in straight lines at unit velocity,
$c \sqrt{\gamma ^2-1} / \gamma $, in the Minkowski spacetime. The
other terms are the first-order approximation for the gravitational
time delay. This is the well-known `Shapiro effect', which has been
amply tested by radar echo delay experiments in our Solar System. Of
course, the total time required for a particle to go form a source
point to an observer can be written as
$t_{OS}=t(D_{OL},r_{0})+t(\sqrt{D_{LS}^{2}+D_{OS}^{2}
\beta^{2}},r_{0})$.

Since particles that form distinct images are almost emitted at the
same time from the same source but travel by different paths, they
may reach an observer at different time. Combining Eq.\,(13) with
the above image equation, we can infer the physical time delay
function between the gravitationally lensed images,
\begin{eqnarray}\label{14}
\Delta t_{im} & = & t_{OS2}-t_{OS1}  \\
 & \simeq &  \frac{r_s}{c} \frac{ \gamma  \left(2 \gamma ^2-1\right) }{2 \left(\gamma ^2-1\right)^{3/2}
}\left[ 2 \bar{\beta}  \sqrt{1+\frac{1}{4} \bar{\beta} ^2} + \frac{2
\gamma ^2-3}{2 \gamma ^2-1} \ln \left(\frac{1+\frac{1}{2}
\bar{\beta} ^2+\bar{\beta} \sqrt{1+\frac{1}{4}\bar{\beta}
^2}}{1+\frac{1}{2} \bar{\beta} ^2-\bar{\beta}
\sqrt{1+\frac{1}{4}\bar{\beta} ^2}}\right) \right] \;, \nonumber
\end{eqnarray}
where $\bar{\beta} = \beta / \theta_{E}$ is the reduced misalignment
angle. Furthermore, we can give the time delay of a relativistic
particle relative to a photon emitted by the same source at the same
time, $\Delta t_{ph}$, which can be measured very easy. As space is
limited, we shall not go into details of the formula.

Besides the multiplicity and distortion of images and the time delay
between different images, another prominent effect of lensing is the
magnifications of the images, which are given by the ratio of the
solid angles of the image and the source,
\begin{equation}\label{15}
\mu_{\pm} = \left|\frac{\theta_{\pm} d \theta_{\pm}}{\beta d \beta}
\right|= \frac{\bar{\beta}^{2}+2}{2 \bar{\beta}
\sqrt{\bar{\beta}^2+4}} \pm \frac{1}{2} \;.
\end{equation}
The measurable total magnification $\mu = \mu_{+} + \mu_{-}$ is
always larger than one. Since the Einstein angle $\theta_{E}$ now is
a slightly larger than that of photon, the image magnification of
neutrino lensing is slightly larger than that of normal lensing. As
usual, the maximal magnification is infinite as $\beta \to 0$,  but
this is not a real situation since the physical objects should have
a finite size $R_{\star}$, the effective limit now is $\beta \to
\beta_{\star} = R_{\star} / D_{OS}$.

Based on 17-year radial velocity and 12-year astrometric
measurements of the short period star S0-2, last year Ghez {\it et
al.}$^{\cite{Ghez}}$ reported the newest and the most precise
results about the central supermassive black hole in our galaxy. The
balk hole is $8.0 \pm 0.6$ kpc from us with mass $(4.1 \pm 0.6)
\times 10 ^{6} M_{\odot}$. Supposed a type of two supernova explode
at 8 kpc away from our galactical center with anggular position
$\beta = 0.1 \sim 0.5 \, \theta_{E}$. The order of magnitude of the
time delay between the primary and secondary images (no matter for
photons or neutrinos) is several dozens seconds. Recently
experimental results (from KamLAND, MINOS, Super-Kamiokande, etc.)
and CMB observation show neutrinos have sub-eV mass (at least one
mass eigenstate state neutrino with a mass of at least $0.04$\,eV,
the best estimate of the mass square difference are $\Delta
m_{21}^2=7.684 \times 10^{-5}$\,eV$^{2}$, $\Delta
m_{32}^2=0.0027$\,eV$^{2}$).$^{\cite{Goobar}, \cite{Yang},
\cite{Amsler}}$ Taken $E_{\nu} \sim 10$\,Mev, the typical value of
supernova neutrinos, $m_{\nu} c^{2} \sim 1$\,eV, the flight time
delay of neutrino relative to photon is about $\sim 10^{-5} \rm{s}
\, \frac{D_{OS}}{16 \rm{kpc}}\sqrt{1+\sin^{2}\beta}$. Thus, we can
really use the forthcoming neutrino facilities to do some more
detailed analysis of time-varying. Combined with the observation of
the normal lensing of light ray, neutrino lensing probably can
provide us some valuable hints about the spacetime characteristic of
the lens object, even the intrinsic quality of neutrino itself.
However we only investigate the point-mass lens model, while the
matter distribution during the propagation may has a strong impact
on the results, so a more detailed and sophisticated treatments are
need in the future.

Acknowledgments: This research was supported by the National Natural
Science Foundation of China under Grants No 10221001.

\end{document}